\documentstyle[11pt,paspconf,epsf]{article}

\begin{document}

\title{Disk galaxy formation and evolution: models up to
intermediate redshifts}

\author{Claudio Firmani and Vladimir Avila-Reese} 
\affil{Instituto de Astronom\'{\i}a, U.N.A.M., Apdo. Postal 70-264 \\
 04510 M\'exico D.F., M\'exico, and}
\affil{Astronomy Department, New Mexico State University \\
 P.O. Box 30001/Dept. 4500, Las Cruces, NM 88003-8001, USA}

\begin{abstract}
Making use of a seminumerical method we develop a scenario of disk galaxy
formation and evolution in the framework of inflationary cold dark
matter (CDM) cosmologies. Within the virializing dark matter halos,
disks in centrifugal equilibrium are built-up and their galactic evolution is 
followed through an approach which considers the gravitational
interactions among the galaxy components, the turbulence and energy
balance of the ISM, the star formation (SF) process due to disk
gravitational instabilities, the stellar evolution and the secular
formation of a bulge. We find that the main properties and
correlations of disk galaxies are determined by the mass, the
hierarchical mass aggregation history and the primordial angular
momentum. The models follow the same trends across the Hubble sequence 
than the observed galaxies. The predicted TF relation is in good
agreement with the observations except for the standart CDM. While the
slope of this relation remains almost constant up to intermediate
redshifts, its zero-point decreases in the H-band and slightly
increases in the B-band. A maximum in the SF rate for most of the
models is attained at $z\sim 1.5-2.5$. 
\end{abstract}

\section{Introduction}

In the last Sesto meeting on galaxy formation and evolution in 1996,
some fundamental questions concerning how do galaxies build up and
evolve in a cosmological frame have been established. Probably one of
the most interesting
question is how do galaxies retain memory about the physics of their
formation, memory that deals with their mass aggregation histories
(MAHs) in a hierarchical cosmogony  and with the origin of their
angular momentum. This
memory may represents a powerful link between the present features of the
galaxies and the nature of the dark matter (DM) as well as the power
spectrum and the statistical properties of its density fluctuation
field (Avila-Reese \& Firmani 1997). The facts that most disk galaxies
have flat rotation curves,
exponential surface brightness profiles, star formation (SF)
histories, colors and gas fractions that correlate with the morphological
type, and show a tight correlation between the
luminosity and the circular velocity, point out to a common physical
origin for many
of their present features. Much progress has been carried out from the
observational and theoretical points of views, particularly since a 
self-consistent scenario for galaxy formation that integrates
cosmology and astronomy has appeared. However, not much
progress has been attained concerning the connection between the
internal physics of galaxies and the overall cosmological frame. 

The current theoretical approaches on galaxy formation can be formulated on the
ground of three main methods: (1) the \textit{numerical simulations},
(2) the \textit{semianalytical} models, and (3) the
\textit{analytical} models. The
numerical simulations (e.g., Yepes 1997 and references therein) are
the most direct way to study the complex
evolution of the density fluctuations, but also, due to technical and 
numerical limitations, at least when including the gas hydrodynamics, this
is currently the less predictive and most expensive method. 
The semianalytical models (Kauffmann, White \& Guiderdoni
1993; Cole et al. 1994; Somerville \& Primack 1998, and more references 
therein) are useful as a method to constrain the 
space of parameters which resume some ignorance about several physical
processes of galaxy formation and evolution, however, they do not
deal with the internal metabolism of the galactic disks. The analytical
method (Mo, Mao \& White 1998; see also Dalcanton, Spergel, \& Summer
1997) is the most economical approach to study the global properties of galaxy 
populations at different redshifts. Nevertheless, this method 
does not deal with the evolution and physics of individual
galaxies. We shall develop a method that allows to calculate the
internal structure and physics of the disk galaxies that forms in the
evolving DM halos, using for this aim simple self-consistent
physical models implemented in a numerical code. The scenario for
which this \textit{seminumerical} method was developed is that of an 
{\it inside-out} galaxy formation where the rate of gas accretion is given
by the rate of the hierarchical mass aggregation, and where the
primordial angular momentum is acquired by the DM halo during its
linear gravitational regime by tidal torques.
\begin{figure}
\vspace{4.0cm}
\includegraphics{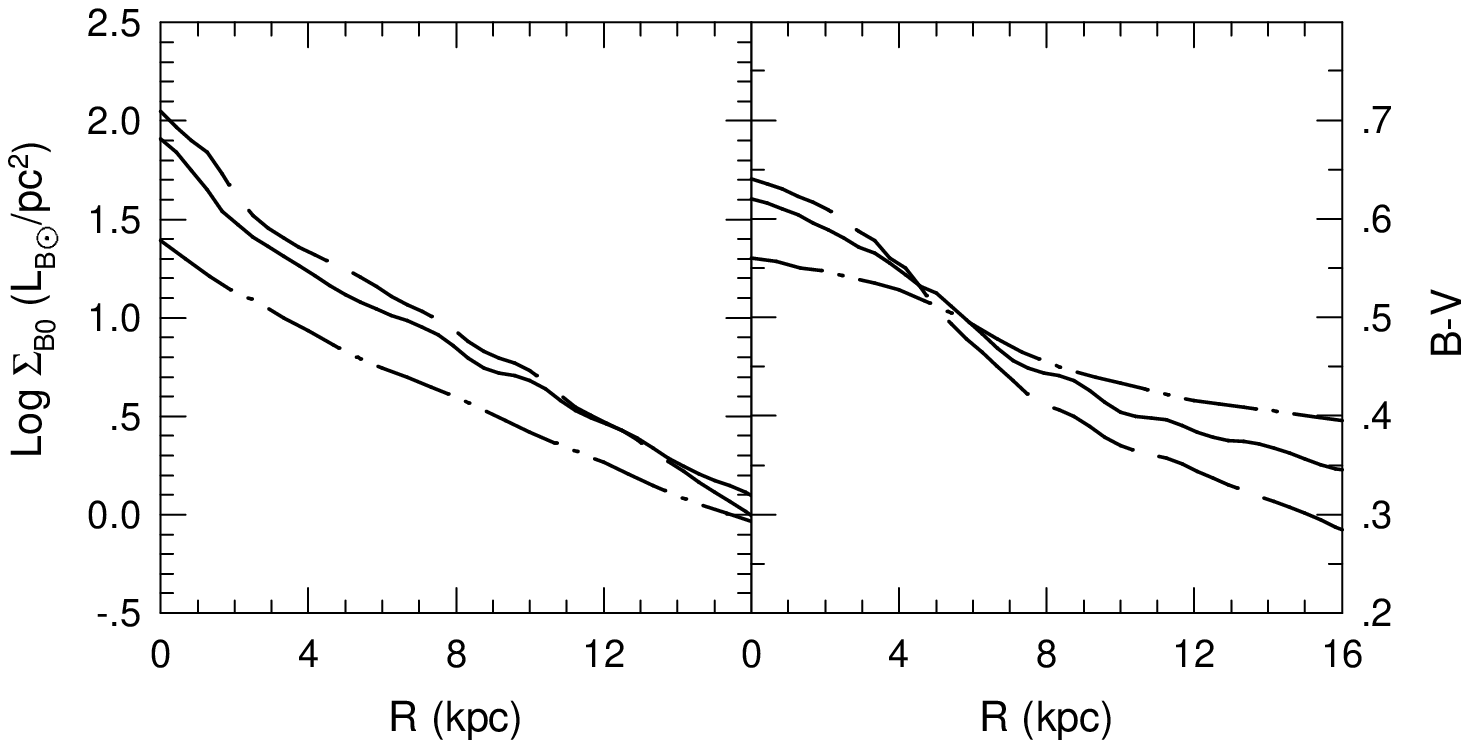}
\caption{The B-surface brightness (left) and the B-V color (right)
profiles of models of 5x$10^{11}M_\odot$ and with the average MAHs and 
$\lambda =0.03$ (dashed line), $\lambda =0.05$ (solid line), and 
$\lambda =0.1$ (point-dashed line). The SCDM, $\sigma_8=0.6$ model was 
used.}
\end{figure}

\section{The physics of galactic disks in a cosmological frame}

\begin{figure}
\vspace{8.7cm}
\includegraphics{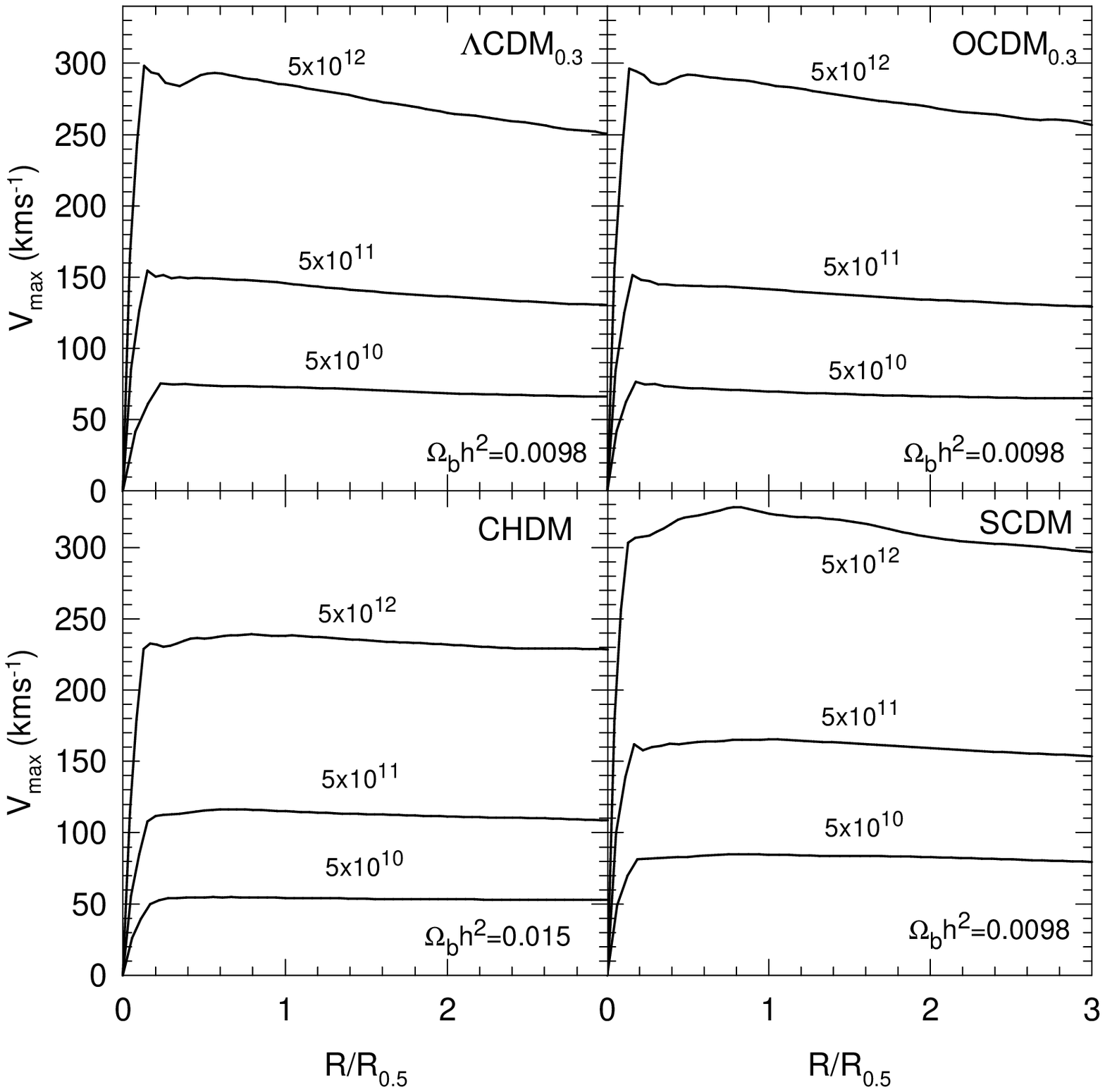}
\caption{The final total rotation curves for models with the average
MAHs, $\lambda =0.05$ and for three masses (shown in the plot). Each 
panel corresponds to the cosmology indicated in the upper right corner.}
\end{figure}

 We start from the hydrodynamics of a gas and a stellar disk with axial
symmetry (Firmani \& Tutukov 1994; Firmani, Hern\'andez \& Gallagher
1996). The disk gas gravitational instabilities induce the SF which is
self-regulated by the SN energy feedback on the ISM. A simple stellar 
population synthesis model is considered to account for the luminosity
evolution. A secular bulge
formation is introduced applying a local gravitational
stability criterion to the stellar disk.  The galactic evolutionary
models are inserted in a cosmological 
background: the structure and evolution of the DM halos which surround
the disks and the mass accretion rates over them are calculated from
initial conditions defined by the cosmological model, which is
specified by the mass fractions of the DM species (cold $\Omega _{CDM}$, 
hot $\Omega _{\nu}$), the vacuum energy ($\Omega _\Lambda $), 
and the baryon matter ($\Omega _b$), and by the value of the Hubble
constant ($h=H_0/100Kms^{-1}Mpc^{-1}$). With the aim to study general 
behaviors we shall use the $SCDM$ model normalized to $\sigma_8=0.6$,
since this model is very frequently found in the literature. In some
cases we also shall present results for some of the most
representative cosmologies: the $\Lambda CDM$ ($\Omega _{CDM}=0.327$, 
$\Omega _\Lambda =0.65$, $h=0.65$), the $OCDM$ ($\Omega _{CDM}=0.327$,
$h=0.65$), the $CHDM$ ($\Omega _{CDM}=0.94$, $\Omega _{\nu}=0.2$, 
$\Omega _b=0.06$, $h=0.5$), and the $SCDM$ ($\Omega _{CDM}=0.96$,
$h=0.5$).  Where the value of $\Omega _b$ is not
specified it was taken equal to $0.01h^2$. A
Gaussian statistical distribution with a power spectrum taken from Sugiyama
(1996) and normalized to the COBE data (except for the $SCDM$
normalized to $\sigma_8=0.6$) characterizes the primordial
density fluctuation field. The MAH of a given halo is obtained by a
Monte-Carlo method
applied on the conditional probability of the density fluctuation
field (Lacey \& Cole 1993). In
this work we are interested in disk galaxies that have not suffered during
their evolution a major merger able to destroy the disk. Then we have
excluded from our sample haloes that in some time have collided with an
other halo with a mass greater than half of its mass at that time. Starting
from the linear evolution of a density fluctuation and assuming  spherical
symmetry we obtain the virialized halo density profile that results from the
non-linear evolution of the density fluctuation. To calculate this we
use a statistical approach based on the adiabatic invariants
(Avila-Reese, Firmani \& Hern\'andez 1998, AFH). The non-radial
component of the kinetic energy is calibrated on the base of the N-body
simulations. The angular momentum of each shell  is
calculated assuming for the virialized structure a  value of the spin
parameter $\lambda $ constant in time (in agreement with the linear theory of
angular momentum acquirement, White 1994). We have
assumed an average value of $\lambda =0.05$ with a lognormal
distribution. Once a mass shell is incorporated into the halo, its
baryon mass fraction falls onto the disk. We assume that
the disk-halo feedback is negligible and that the cooling flow does
retain just a negligible amount of gas within the halo. This is
consistent with the hypothesis that the SN energy feedback is mainly 
localized into the disk in
agreement with the observational evidence. We have used a simple
parametrized gas cooling mechanism into the halo in order to test the
sensitivity of the results to this very complex process. The gas of
the shell is distributed across the disk assuming  rigid rotation of the shell, in
agreement with the Zeldovich approximation, and assuming
detailed angular momentum conservation during the gas dissipative collapse.
Adiabatic invariants are used to calculate the halo gravitational
contraction produced by the increasing gravitational field due to the disk
mass growth.

\begin{figure}
\vspace*{5.4cm}
\includegraphics{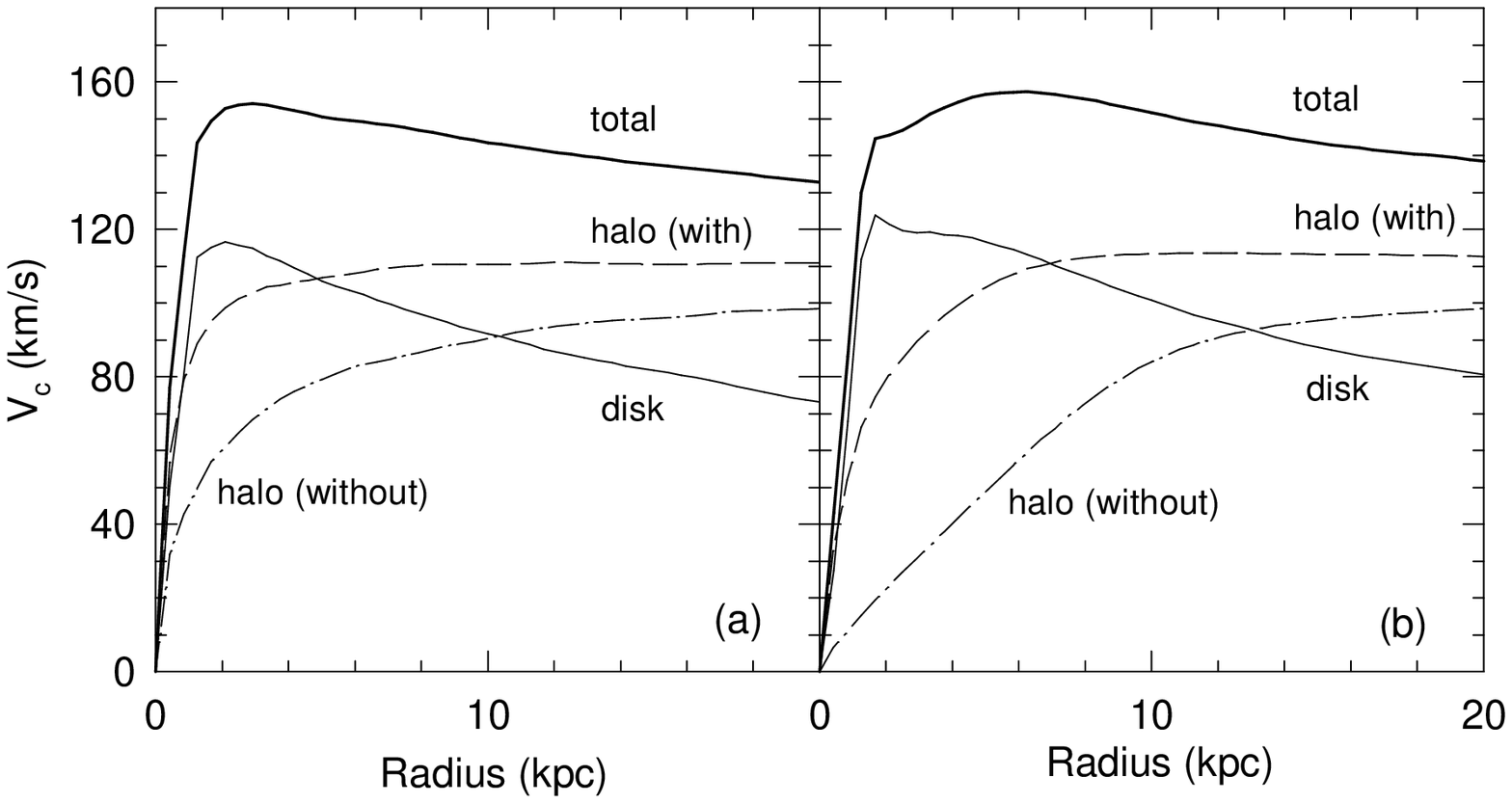}
\caption{Rotation curve decompositions ($\Lambda CDM$ model). The DM halo 
contribution {\it with} and {\it without} the gravitational
contraction due to the disk
is shown. Panel (b) is for a DM halo with a near constant density core.}
\end{figure}

\section{Results}
For a fixed present-day mass we obtain a large sample of possible MAHs 
(see fig. 1 in Firmani \& Avila-Reese 1998) which after the
virialization give rise to a {\it wide range} of density profiles. The 
average MAH produces a density profile similar to that obtained in
N-body simulations by Navarro, Frenk and White
(1997). However, our method gives in a natural way the statistical
deviations from the average MAH, which lead to a {\it rich variety
of halo structures} (AFH). The properties of luminous
galaxies may depend on the features of these evolving halos because of 
luminous galaxies form within them. 

\subsection{Local properties}
The surface density profiles of the disks in centrifugal equilibrium formed
within the DM halos with the gradually infalling gas shells in solid body
rotation are nearly exponential over several scale lengths
(fig. 1a). Models with high angular momentum have extended disks with
low surface brightness. According to Dalcanton, Summer \& Spergel
(1997), these models would correspond to the low surface brightness
(LSB) galaxies. Observations show systematic blue
radial color gradients in spiral galaxies (e.g., de Jong 1996a). Our
models show a similar trend (fig. 1b), although some excessive
gradient appears compared to the observations. This problem may be due
to angular momentum redistribution during the gas infalling or to some 
inaccuracy of the simple population synthesis technique that works in
our code. The former one is particularly a very complex problem that
we shall explore in future works. 

\begin{figure}
\vspace*{5.1cm}
\includegraphics{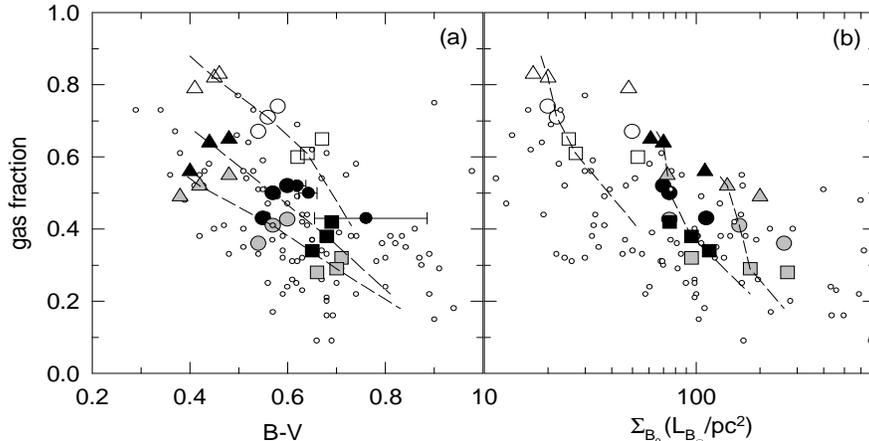}
\caption{The gas fraction $f_g$ vs. the B-V color 
(a), and vs. the central B-surface brightness $\mu _{B_0}$ (b) for
  models and observations. The gray, black, and white filled symbols,
  correspond to models with $\lambda =0.035,$ $\lambda =0.050,$ and
  $\lambda =0.100$ respectively. Squares are for the early active MAH,
  circles for the average
MAH, and triangles for the extended MAH. Three masses (dark+baryon), 5$%
\times 10^{10}M_{\odot },$ 5$\times 10^{11}M_{\odot },$ and 5$\times
10^{12}M_{\odot }$ are considered (the larger the mass, the smaller is
the
gas fraction). The dashed lines connect the models of constant mass for 5$%
\times 10^{11}M_{\odot },$ and extend the statistical range of MAHs to
94$\%$ (symbols consider only 80$\%$ of the MAHs). The three small black
filled circles are the same models corresponding to the big black
filled circles but reddened according to the dust
absorption-luminosity dependence given in Wang \& Heckman (1996) (see
text). The error bars correspond to the range of values which fit
observational data. Small empty circles are the observational
data. LSB galaxies are included.}
\end{figure}

\begin{figure}
\vspace*{5.1cm}
\includegraphics{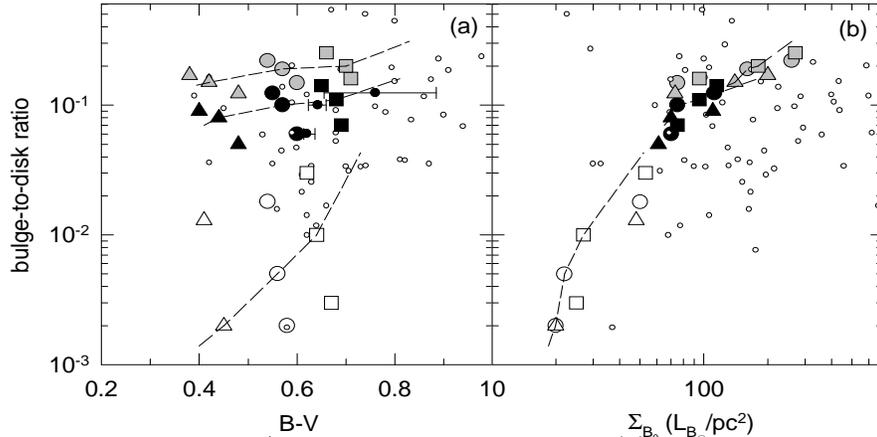}
\caption{The b/d ratio vs. the B-V color
  (a), and the central B-surface brightness $\mu _{B_0}$ (b) for
  models and observations. The same symbol and line codes of fig. 4
  are used. LSB galaxies and a few normal galaxies shown in fig. 4 are
absent in this fig.}
\end{figure}

The rotation curves obtained for the average
MAH and $\lambda =0.05$
for the masses $5\times 10^{10}M_{\odot }$, $5\times 10^{11}M_{\odot}$
and $5\times 10^{12}M_{\odot }$ and for the cosmological models taken into
account here are nearly flat (fig. 2). This explains the cosmological
nature of the conspiracy between baryon and dark matter in the flat profile of
the rotation curves. The decomposition of the
rotation curve for the model with $5\times 10^{11}M_{\odot }$ 
($\Lambda CDM$) is presented in fig.3a, where the halo gravitational
contraction due to the
disk is shown comparing the halo component \textit{with} and 
\textit{without} this contraction. We note here the problem already
pointed out in Burkert (1995) (see also Flores \& Primack 1994, Moore
1994): the gravitational contribution of the halo is dominant until
the center. Even if the observational methods to decompose a rotation
curve are affected by a large
uncertainty, some physical process may be being misundertood
here. Recent high-resolution numerical results obtained by Kravtsov et
al. (1998) might be showing that the problem is not too serious,
and, comparing with our results, suggest that the merging process and
the slope of the power spectrum at the scale in consideration are
the responsible of giving rise to shallow cores in the DM halos. If we
artificially shallow the central regions of our predicted density
profiles in order to be in agreement with the rotation curves of the dwarf
and LSB galaxies, then the final rotation curve
decomposition results more realistic (fig.3b).

\begin{figure}
\vspace*{5.3cm}
\includegraphics{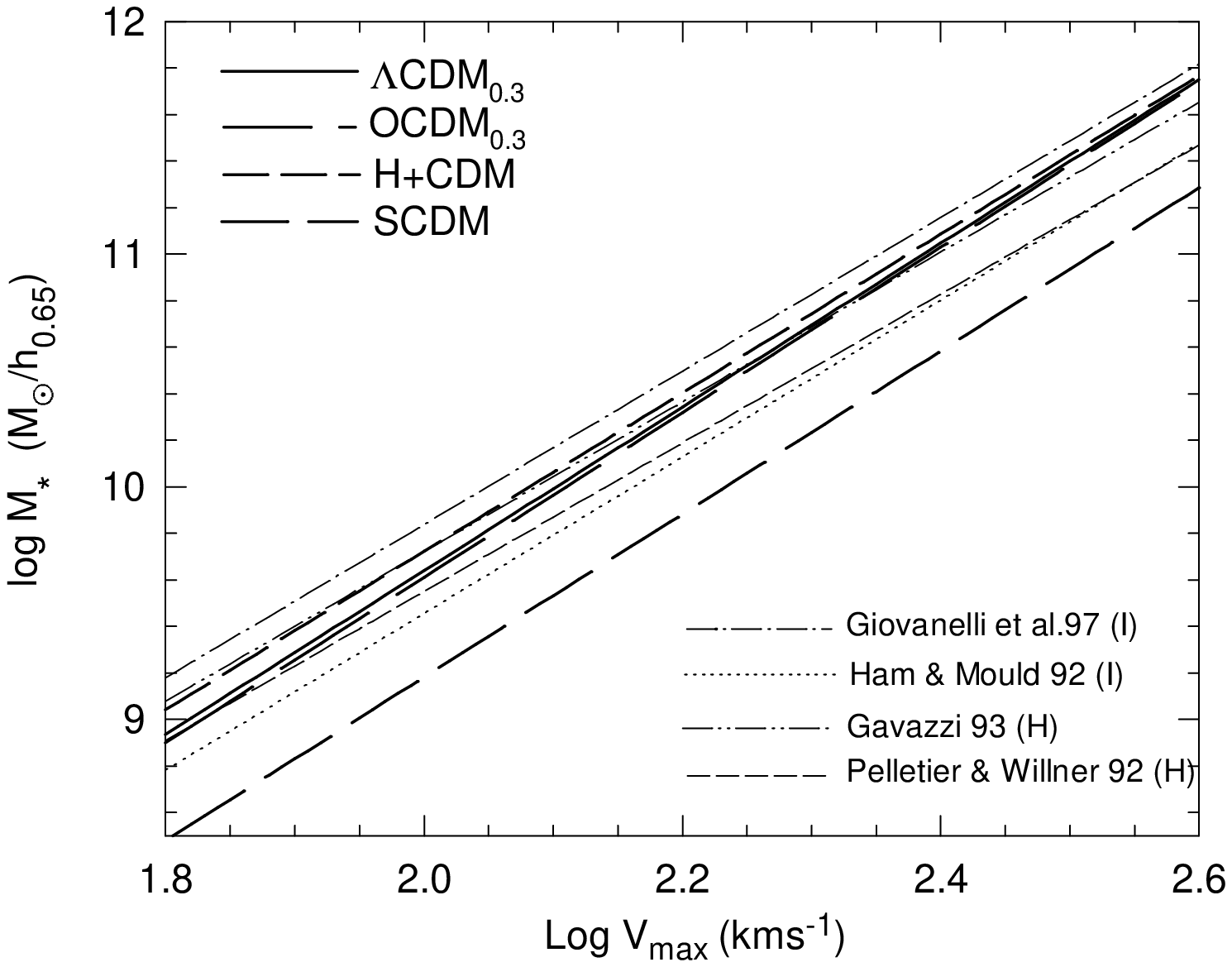}
\caption{The predicted TF relation for several cosmologies compared
with the observational data in the I- and H-bands. For the references see
AFH.}
\end{figure}

\subsection{Global properties}

 In fig. 4 and 5 we compare some global properties obtained for the
$SCDM$ model normalized to $\sigma_8=0.6$ (for the other models the
situation is nearly the same) with the observations. In
fig. 4a and 4b the gas fraction is
displayed \textit{vs.} $B-V$ color and the central surface brightness,
respectively. The data were taken from a compilation given in McGaugh
and de Blok (1996) and include LSB galaxies. In fig.
5a and 5b the disk-to-bulge (b/d) ratio is displayed \textit{vs.} the same
variables of fig. 4. It is surprising that the b/d ratios predicted by 
the models using a simple gravitational instability criterion are in
agreement with those inferred from observations (de Jong 1996b). Our
models reproduce the observational trends across the Hubble sequence:
{\it the redder and more concentrated the disks, the smaller are the gas
fractions, and the larger are the b/d ratios}. The models seem to
follow a biparametrical sequence, where the two parameters are the 
color index B-V and the central surface brightness. These two properties
are mainly determined by the MAH and by the value of $\lambda $,
respectively. The intensive global properties are almost independent
on the third initial fundamental factor, the mass, which is important
in defining the extensive properties (luminosity, radius, etc.).

\subsection{The Tully-Fisher relation}

In fig. 6 we plot the stellar disk mass $M_{*}$ {\it vs.} $V_{\max
}$ obtained from our simulations for the different cosmological
models. The observed I-band TF relation was transformed to the 
$M_{*}-V_{\max }$ relation using the
mass-to-luminosity radio $\Upsilon _I=1.8*(\frac{M_{*}}{5*10^10M_{\odot }}%
)^{0.07}h$ (AFH98).  For the H-band TFR we have used $\Upsilon _H=0.5h;$ this
mass-to-luminosity ratio is obtained from direct observational estimates in
the solar neighborhood (Thronson \& Greenhouse 1989). The agreement in
the slope as well as in the zero-point (except for the COBE-normalized
SCDM model) between the theoretical (cosmological) and
empirical $M_{*}-V_{\max }$ relations is excellent, pointing out this to a
{\it cosmological origin} for the I- or H- band TF relation. The TF
relation represents a fossil of the primordial density fluctuation field. 

This structural relation of
galaxies is mainly imprinted by the power spectrum of fluctuations, that for
the CDM models at galactic scales produces a variance rather independent
on mass, which implies virialized objects whose average densities are almost
independent on mass $<\rho >\propto M^{-\alpha}$, $\alpha <<1$, or,
that is the same, using $V_c^2\propto M/R$,
whose masses scale as $\propto V_c^{6/(2-\alpha)}$, i.e. since $\alpha
<<1$, approximately $M\propto V_c^3$. The mass-velocity relation for
the DM haloes indeed scales as the velocity to 3.3-3.4 (AFH; see
also Navarro et al. 1997); this relation remains almost unaltered for
the luminous disks which form within the DM halos. We have explored the
robustness of this result with respect to the baryon fraction and the infall
dissipative processes that retain the gas in the halo. In both cases we find
that the TF relation is almost not affected, mainly due to the
conspiracy of the gravitational contraction of the dark
halo: if the fraction of mass which goes to the disk decreases
(increases), then
the final rotation velocity also decreases (increases) in such a way that the
mass-velocity relation remains nearly the same one. Different
approaches based on numerical simulations arrive to similar
conclusions (Elizondo et al. 1998). 
 
 Regarding the scatter of the TF relation we find that it is produced
by the dispersion of the MAHs for a given mass and by the statistical
distribution of the spin parameter $\lambda$. As an extreme case we
assume that they are independent one from another. For the $\Lambda
CDM$, $OCDM$ and $CHDM$ models we predict a total scatter of 
0.4-0.5 mag that is in marginal agreement with the observations. It is 
important to have in mind that in our models, at difference of the
observational works, the LSB galaxies are
included. For the $SCDM$ models (COBE or $\sigma_8=0.6$ normalized)
the scatter is rather high, larger than
0.6 mag. Our models show that the scatter of the TF relation
correlates with some galaxy properties. In fig. 7 it is shown how for
a given mass the maximum rotation velocity does correlate with the B-V
color of the galaxy. The observational data seem to follow the same
trend.

\begin{figure}
\vspace*{6.9cm}
\includegraphics{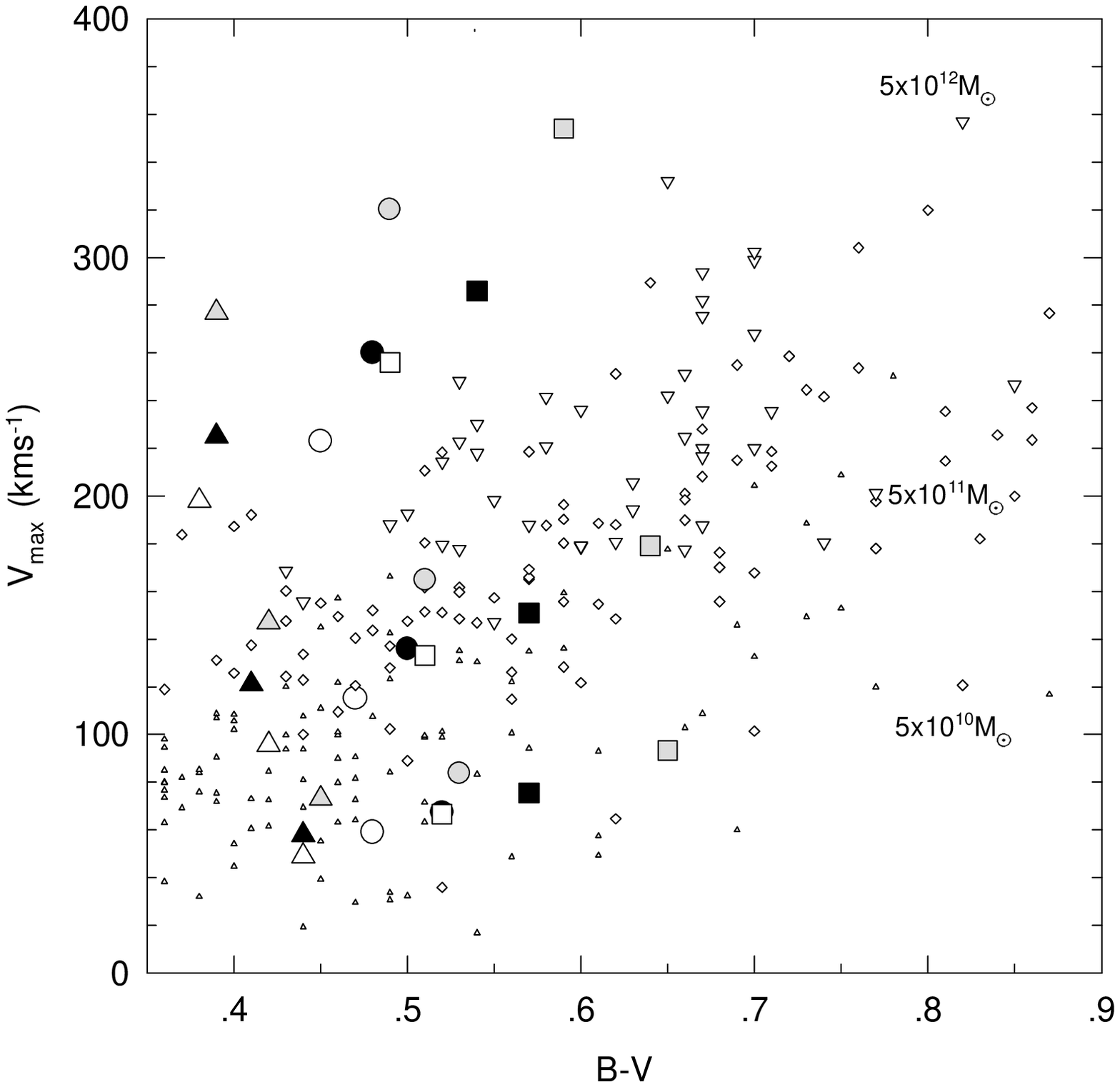}
\caption{The maximum rotation velocity vs. B-V for models and
  observations. The same symbol codes of fig. 4 are used. The
  observational data (small symbols) were taken from a cross of the
  RC3 and the Tully (1988) catalogs. The small triangles,
  diamonds, and inverse triangles
correspond to galaxies with luminosities in B-band within the 10$^8-3%
\times ${10}$^9L_{B_{\odot }},$ 3$\times ${10}$^9-3\times
10^{10}L_{B_{\odot }}$, and 3$\times 10^{10}-2\times
10^{11}L_{B_{\odot }}$ ranges, respectively. Note how the maximum
velocity of models and observations for a
given mass (or range of luminosities) correlates with the B-V color.}
\end{figure}

 In the B-band we obtain a TF relation with slopes roughly of 3.5 for
the different cosmological models. The observational data give
slopes between 2 and 3. This discrepancy may be due to the extinction
in galaxies and its dependence with mass. Wang \& Heckman (1996), for
a large sample of galaxies, have found that the dust opacity increases
with the luminosity. Using the opacity-B luminosity relation they give 
we have ``decorrected'' the B-band luminosities of our models. After
this the slope of the TF relation in the range $\sim
10^9-10^{11}L_{B_{\odot }}$ is well approximated by a line with slope 
$\sim $2.7, i.e. the predicted TF relation now agrees with the
observational estimates (see fig. 8). 
\begin{figure}
\vspace*{5.8cm}
\includegraphics{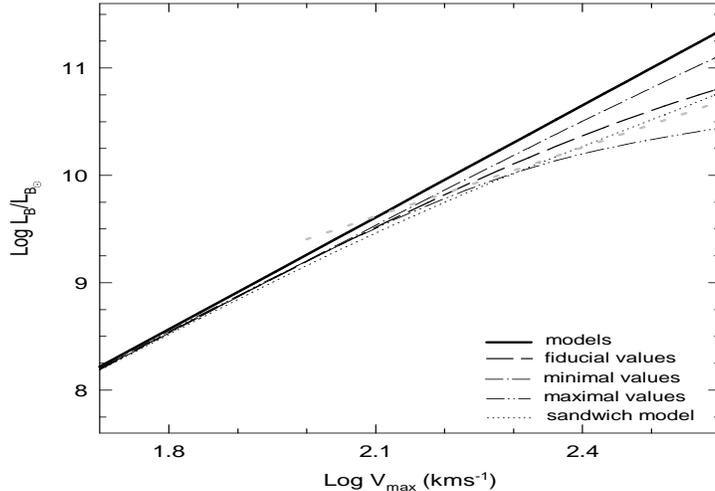}
\caption{The predicted B-band TF relation for the $SCDM$, $\sigma _8=0.6$
 model (thick solid line).
  The other lines show how the intrinsical TF relation transforms if
  the B-luminosities are dimished by dust absorption according to the
  observational dependence of optical depth of dust on luminosity
  given in Wang \& Heckman (1996). The dashed line corresponds
  to the fiducial optical depth. In the range $10^9-10^{11}L_{B_{\odot }}$
  this line is well approximated by a line with slope $\sim
  2.7$. The dotted gray curve is the linear regresion to the data
  given in Kudrya et al. (1997). We have truncated the
  regresion at $V_{\max }=100$ km/s because it does not provide a good
  approximation for lower velocities (see Figure 6 of Kudrya et al.
  1997).}
\end{figure}

The seminumerical approach allows us to follow the
evolution of an individual
galaxy, then we are able to predict how a given galaxy does appear at different
redshifts. Fig. 9 shows the behavior of $V_{\max }$vs. $z$. Here we see the
difficulty of $CHDM$ to produce galaxies with high rotation velocity in the
redshift range proper of the damped $Ly\alpha $ absorbers (e.g., Klypin
et al. 1995). In fig. 10 we show our predictions for the evolution of
the TF zero-point in
the H and B bands (the slopes remain almost constant). The
zero-point of the H-band TF relation decreases with $z$ because
galaxies after reaching their maximum rotation velocity continue
aggregating matter without changing this velocity. In the B-band,
because the B-luminosity increases with $z$, some compensation
happens in such a way that the zero-point of the TF relation remains
almost constant as deep field observations have recently shown (Vogt
et al. (1997); see fig. 10). Fig. 11 displays the disk scale radius
{\it vs.} (1+z). An inside-out evolution is evident.
\begin{figure}
\vspace*{4.9cm}
\includegraphics{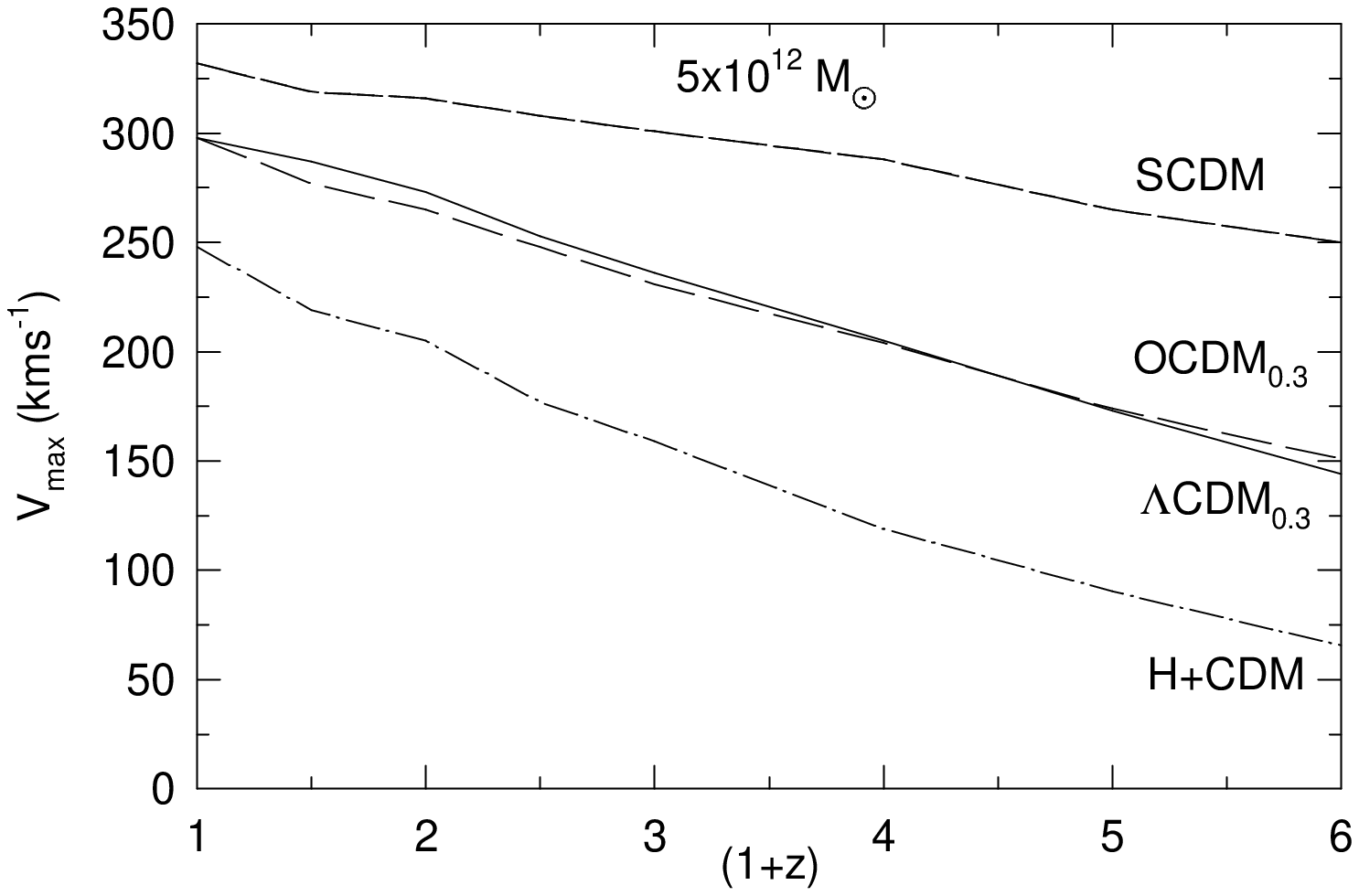}
\caption{The maximum rotation velocity {\it vs.} (1+z) for different 
cosmologies.}
\end{figure}

\begin{figure}
\vspace*{4.5cm}
\includegraphics{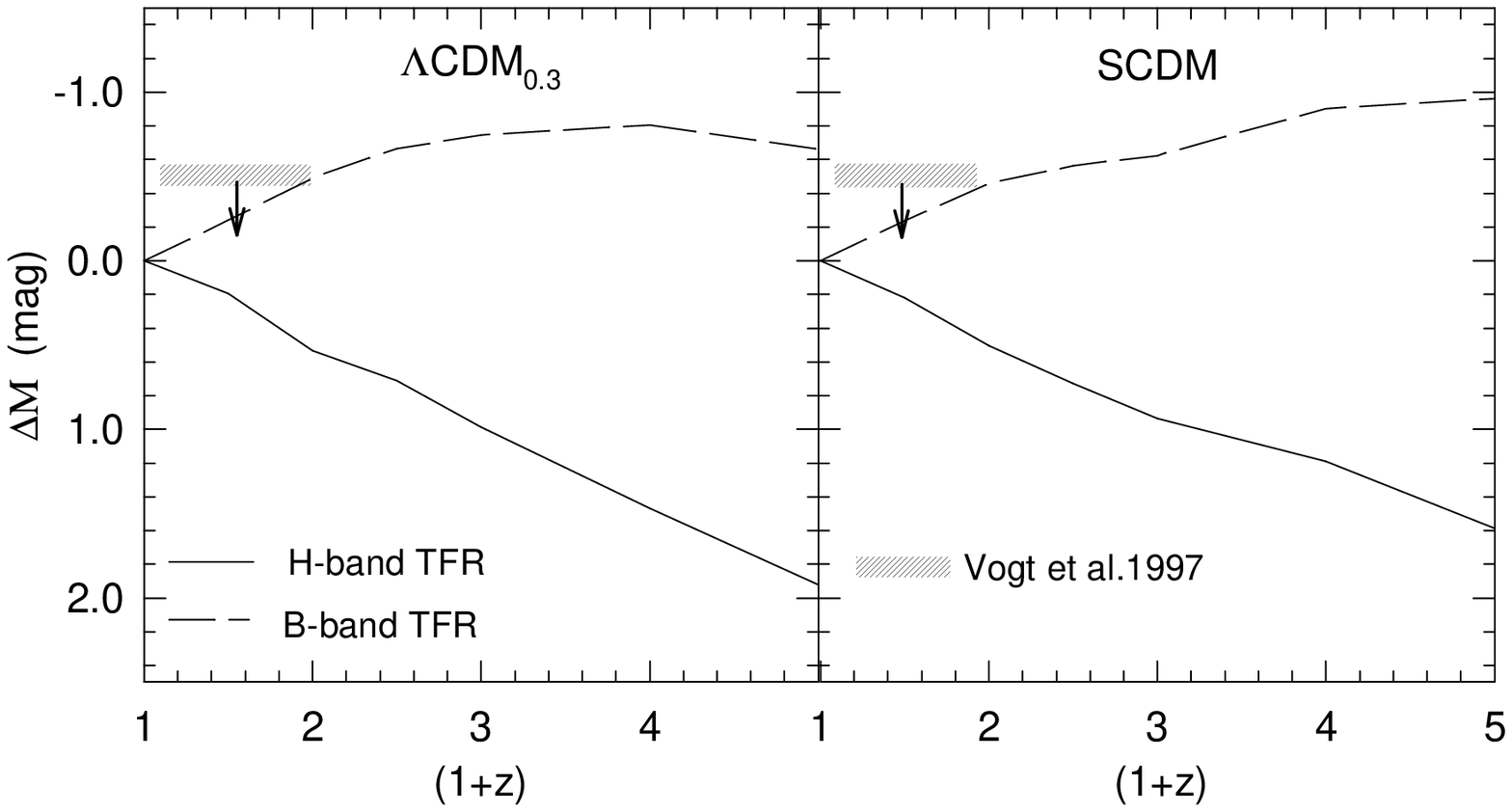}
\caption{The evolution of the TF zero-point in the H- and B-bands. The 
slope of this relation remains almost constant up to high
redshifts. The zero-point in the H-band increases with $z$ because of the mass
(luminosity) of the galaxies continues increasing while their maximum
circular velocities (compactness) remain nearly constant since high redshifts.
In the B-band this effect is compensated, and even reversed because
the B-luminosity increases to the past (the SF rate has its maximum at 
redshifts 1.5-2.5). The shadow bar is an upper limit in the B-band for $z<1$
obtained from deep field observations (Vogt et al. 1997).}
\end{figure}

 Regarding the SF history we find that it is controlled by the gas
accretion (determined by the MAH) and by the disk surface density
(determined by $\lambda $). For the average MAHs and $\lambda 's$ the
SF rates in the
$\Lambda CDM$ and $OCDM$ cosmological models attain a maximum at $z\sim
1.5-2.5$ depending on the mass.  This rate is approximately 2-3
times larger than the present one, a result that is in 
marginal agreement with deep field observations of a selected
population of large spiral galaxies (Lilly et al. 1998). Models with low
$\lambda 's$ and/or violent MAHs attain  maximum SF rates that are
3-6 times larger than at the present epoch. The $SCDM$
models show an earlier and higher peak in the SF rate. In the case of
the $CHDM$ model the SF rate peaks at redshifts smaller than 0.5 and
it remains almost constant up to $z=0$. It is important to remark
that our evolutionary tracks concern
basically to isolated galaxies, where the environment may supply any amount of
gas to a galaxy according to its gravitational field. Our results are not
representative of the average conditions of the universe neither of more
complex situations as the case of galaxies in clusters. More work is planned
in the future on this direction.

\begin{figure}
\vspace*{5.9cm}
\includegraphics{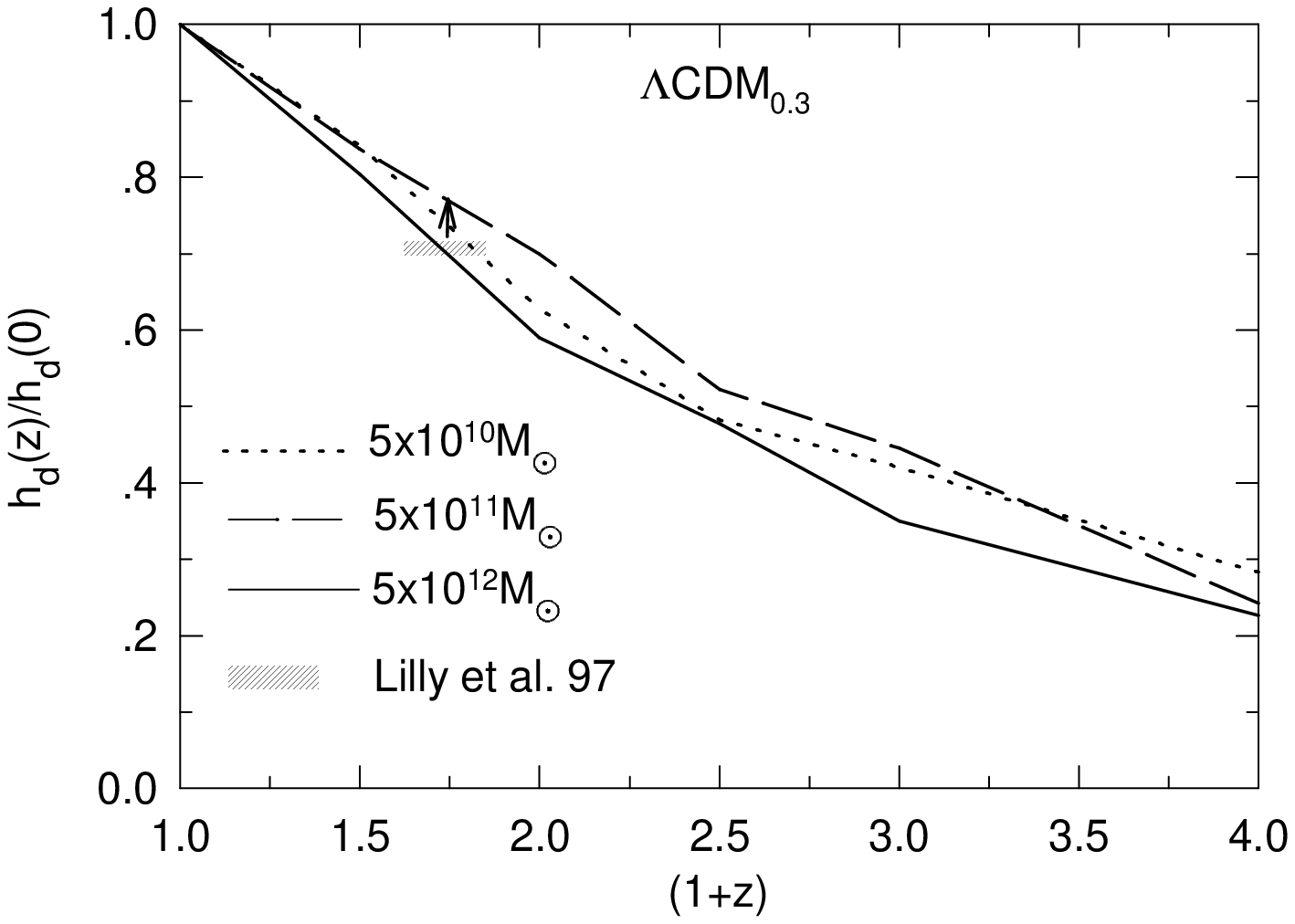}
\caption{The scale radius {\it vs.} (1+z) and a lower limit established by the
observations.}
\end{figure}

\section{Conclusions}

1) The \textit{seminumerical} models presented in this work support the
viability of an inside-out disk galaxy formation scenario, where the rate of
gas accretion to the disk is dictated mainly by the cosmological
(hierarchical) mass aggregation rate, and where the SF occurs in a stationary
self-regulated regime {\it within} the disk.  2) The properties,
correlations and evolution features of the models are tightly related
to three main factors determined by the initial cosmological
conditions: the MAH, the angular momentum and the mass. 3) The
infrared TF relation manifest itself as a clear
imprint of the power spectrum of fluctuations. The
gravitational pull of the luminous matter on the dark halo makes this
relation robust with respect to intermediate processes (cooling,
feedback). In the B-band the luminosity dependent extinction gives
rises to a decrement of the TF relation slope; this may explains the
``color'' TF relation of disk
galaxies. 4) The scatter of the TF relation for most of the models is
in marginal agreement with those derived from the observations. 5) The
slopes 
of the TF relations do not change up to
intermediate redshifts. While in the H-band the zero-point decreases,
in the B-band it remains almost constant. 
6) Concerning the predictive abilities of different cosmologies with respect
to galaxy formation and evolution: i) $\Lambda CDM$, $OCDM$ cosmologies are
able to predict many of the galaxy features up to intermediate redshift. ii) 
$SCDM$ is ruled out because is unable to predict the Tully-Fisher
relation, and gives a too high scatter for this relation. 
iii) $CHDM$ is marginal because predicts a too late galaxy formation
and SF history.

\end{document}